# New methodology based on energy flux similarity for large-eddy simulation of transitional and turbulent flows

Han Qi[1,2], Xinliang Li[1,2], Hao Zhou[1,2], and Changping Yu[1]†

[1]LHD, Institute of Mechanics, Chinese Academy of Sciences, Beijing 100190, China

[2]School of Engineering Science, University of Chinese Academy of Sciences, Beijing 100049, China

A new methodology based on energy flux similarity is suggested in this paper for large eddy simulation (LES) of transitional and turbulent flows. Existing knowledge reveals that the energy cascade generally exists in transitional and turbulent flows with different distributions, and the characteristic quantity of scale interaction in energy cascade processes is energy flux. Therefore, energy flux similarity is selected as the basic criterion to secure the flow field getting from LES highly similar to the real flow field. Through a priori tests, we find that the energy flux from the tensor-diffusivity (TD) model has high similarity with the real energy flux. Then, we modify the modelled energy flux from the TD model and obtain uniform formulas of energy flux similarity corresponding to different filter widths and locations in the wall-bounded turbulence. To secure the robustness of simulation and the LES results similar to the real flow, we apply the energy flux similarity method (EFSM) to the Smagorinsky model in the LES of compressible turbulent channel flow, compressible flat-plate flow, and flow over a compressible ramp. The a posteriori tests show that, overall, EFSM can better predict these flows than other subgrid-scale models. In the simulation of turbulent channel flow, EFSM can accurately predict the mean stream-wise velocity, Reynolds stress, and affluent coherent structures. In LES of compressible flat-plate flow, EFSM could provide accurate simulation results of the onset of transition and transition peak, skin friction, and mean stream-wise velocity in cases with three different grid scales. Meanwhile, for flow over a compressible ramp, EFSM could correctly describe the process of bypass transition, locations of separation and reattachment in the corner region, and abundant coherent vortex structures, etc. All the analysis results show that EFSM can efficiently solve several classical difficulties in LES including those for compressible flows, transitional flows, and separated flows, etc. Overall, EFSM is a scale-adaptive method and does not require test filtering and wall model which is suitable for LES of practical wall-bounded flow with complex geometric boundary.

## 1. Introduction

Thus far, large-eddy simulation (LES) has achieved great success in numerical simulation of turbulent flows, and it has already been widely used in studying the flow mechanism of large Reynolds number turbulence and simulating some practical engineering flows (Larchevˆeque et al. 2004; Fureby 2008). In traditional LES, the most widely used subgrid-scale (SGS) model is the eddy-viscosity model, and the first SGS eddyviscosity model is the Smagorinsky model proposed by Smagorinsky (1963) and Deardorf (1970). Using the eddy-damped quasi-normal Markovian theory, Chollet & Lesieur (1981) suggested the spectral eddy-viscosity model, which is suitable for homogeneous and isotropic turbulence. Based on the square of the velocity gradient tensor,

Nicoud & Ducros (1999) proposed a new SGS model named wall-adapting local eddy-viscosity model (WALE), which can simulate wall-bounded flows without the dynamic procedure. The Vreman (2004) model is another SGS eddy-viscosity model suitable for LES of turbulent shear flflows. Recently, Yu et al. (2013) presented a new form of SGS viscosity according to SGS helicity dissipation balance and a spectral relative helicity relation in the inertial subrange of helical turbulence, and this model can simulate the shear and separated turbulent flows with satisfactory results. In addition to the eddy-viscosity model, the structural model is also an important type of SGS model, which provides higher correlation with the real SGS stress. Clark et al. (1979) and Vreman et al. (1996) employed different expansion methods to obtain the gradient model (GM) respectively. The SGS stress can also be modelled as a scale similarity model (SSM) on the basis of the scale similarity hypothesis (Bardina et al. 1980; Liu et al. 1994). In addition, some LES methods have also been developed to promote the SGS models to simulate turbulent flows more precisely. The famous dynamic procedure was proposed by Germano et al. (1991), using the Germano identity to determine the coefficient of the SGS model dynamically in LES of turbulent flows. Subsequently, Lilly (1992), Piomelli (1993), and Meneveau et al. (1996) improved and generalized the dynamic procedure, promoting the dynamic procedure to become the most commonly used method in LES of turbulence. With the Germano identity, Yu et al. (2016) derived an expression of the energy flux at the test-filter scale, which can be adopted to optimize the coefficient of SGS models. Recently, Chen et al. (2012) introduced Reynolds stress to constrain the SGS model in the near-wall region of wall-bounded turbulent flows. With the constrained LES method, statistical average results near the wall could be improved apparently. Nevertheless, the traditional LES still faces several challenges, such as the simulation of transitional flows (Sayadi & Moin 2012), application to compressible flows (Piomelli 1999), and limitation of the scale-invariance hypothesis (Voke 1996; Meneveau & Katz 2000). In order to solve these problems, several SGS models and LES methods have been developed recently. Transition to turbulence is important to flow mechanism and engineering research, but the prediction of the transition remains difficult in LES. Horiuti (1986) employed the standard Smagorinsky model to simulate a transitional channel flow and found that this SGS model cannot predict the transition process due to excessive dissipation. Huai et al. (1997) applied the dynamic Smagorinsky model (DSM) to the simulation of a transitional flflat-plate boundary layer for the first time, and obtained acceptable results. Sayadi & Moin (2012) evaluated several commonly used SGS models and methods in LES of transitional flows. By analysing the simulation results, they found that the dynamic procedure could predict the transition, but the results of some constant coefficient SGS models were invalid in the simulation of turbulence. Recently, Bodart & Larsson (2012) added a laminar/turbulent sensor to the traditional wall model and successfully predicted the transition. In recent years, LES of compressible turbulent flows has been attracting increasingly more attention. However, SGS models or LES methods developed for LES of compressible turbulent flows are still lacking. Moin et al. (1991) suggested the compressible DSM model for the first time and applied the model to the simulation of compressible isotropic turbulence. Chai & Mahesh (2012) proposed a dynamic one-equation eddy viscosity model for LES of compressible flow and applied it to decaying isotropic turbulence and normal shock-isotropic turbulence interaction. Xu et al. (2010) used DSM to simulate compressible flow past a wave cylinder and studied the mechanism of the flow.

In traditional LES, most SGS models and LES methods are based on the hypothesis of scale invariance, and the grid scale needs to be in the inertial subrange. However, the grid scale could not always located in the inertial subrange actually. Voke (1996) analysed several hypothesized full-range energy spectra (e.g., Heisenberg-Chandrasekhar spectrum; Kovasznay spectrum; Pao spectrum), and provided a fitted relation between SGS viscosity of the Smagorinsky model and mesh Reynolds number. This scale-dependent Smagorinsky model is an attempt to overcome the limitation of the scale-invariance hypothesis, and the simulating results were slightly improved when the cutoff was in the dissipation range. Using the bi-dynamic procedure in an a priori test, Meneveau & Lund (1997) supplied a fitting ratio of the test-scale to grid-scale coefficient of the Smagorinsky model, and applied the scale-dependent dynamic Smagorinsky model to LES of forced isotropic turbulence. Port´e-Agel et al. (2000) then generalized the scaledependent dynamic Smagorinsky model to LES of a neutral atmospheric boundary layer. On the basis of the Kovasznay spectrum, Yu et al. (2017) deduced the expression of the ratio between SGS dissipation and resolved viscous dissipation at an arbitrary grid scale. This scale-adaptive LES method can be easily implemented in single and mixed models for LES of isotropic and wall turbulence. In this paper, we propose a new LES method based on energy flux similarity in an attempt to solve the current challenges of LES. The structure of the paper is as follows: the LES governing equations and modelling theoretical background are introduced in §2. Energy flux similar method are proposed in §3, followed by a posteriori tests in §4, where the LES results of turbulent channel flow, transition and turbulent boundary layer and turbulent boundary layer over a compression ramp are presented. Finally, the discussion and conclusions are given in §5.

† Email address for correspondence: cpyu@imech.ac.cn

## 2. Theoretical background

### 2.1. LES Governing Equations

For the general applicability of the research, we select LES Governing Equations of compressible flows as follows

$$\frac{\partial \overline{\rho}}{\partial t} + \frac{\partial \overline{\rho}\widetilde{u}_j}{\partial x_j} = 0, \tag{2.1}$$

$$\frac{\partial \overline{\rho}\widetilde{u}_i}{\partial t} + \frac{\partial \overline{\rho}\widetilde{u}_i\widetilde{u}_j}{\partial x_j} + \frac{\partial \overline{p}}{\partial x_i} = \frac{\partial \widetilde{\sigma}_{ij}}{\partial x_j} - \frac{\partial \tau_{ij}}{\partial x_j}, \tag{2.2}$$

$$\frac{\partial \overline{\rho}\widetilde{E}}{\partial t} + \frac{\partial \left(\overline{\rho}\widetilde{E} + \overline{p}\right)\widetilde{u}_j}{\partial x_j} = -\frac{\partial \widetilde{q}_j}{\partial x_j} + \frac{\partial \widetilde{\sigma}_{ij}\widetilde{u}_i}{\partial x_j} - \frac{\partial C_p Q_j}{\partial x_j} - \frac{\partial J_j}{\partial x_j}, \tag{2.3}$$

$$\overline{p} = \overline{\rho}R\widetilde{T}, \tag{2.4}$$

where a bar denotes spatial filtering at scale $\Delta$ using a smooth low-pass filter function $G_\Delta(r)$ (e.g., $\overline{\rho}(x) = \int G_\Delta(\boldsymbol{r})\rho(x+\boldsymbol{r})d\boldsymbol{r}$ ) represents the resolved density field) and a tilde denotes spatial Favre filtering as $\widetilde{\phi} = \dfrac{\overline{\rho\phi}}{\overline{\rho}}$.

In (2.1) - (2.4), $\rho$, $u_i$, $T$, $E$ and $R$ denote density, velocity, temperature, total energy, and specific gas constant, respectively. The viscous stress tensor and the heat flux vector are given by

$$\widetilde{\sigma}_{ij} = 2\mu\left(\widetilde{T}\right)\left(\widetilde{S}_{ij} - \frac{1}{3}\delta_{ij}\widetilde{S}_{kk}\right), \tag{2.5}$$

† Email address for correspondence: cpyu@imech.ac.cn

$$\widetilde{q}_j = \frac{C_p \mu\left(\widetilde{T}\right)}{P_r} \frac{\partial \widetilde{T}}{\partial x_j}, \tag{2.6}$$

where $C_p$, $P_r$ are the specific heat at constant pressure and molecular Prandtl number, $\mu = \frac{1}{\mathrm{Re}} \frac{\widetilde{T}^{3/2}\left(1 + T_s / \overline{T}_\infty\right)}{\widetilde{T} + T_s / \overline{T}_\infty}$ is the molecular viscosity calculated using Sutherland's law for given $Ts = 110.3K$, $Re = \rho_\infty U_\infty L / \mu_\infty$ is the Reynolds number, and $\widetilde{S}_{ij} = \frac{1}{2}\left(\dfrac{\partial \widetilde{u}_i}{\partial x_j} + \dfrac{\partial \widetilde{u}_j}{\partial x_i}\right)$ is the resolved strain-rate tensor.

In (2.1) - (2.3), there are some unclosed terms, the SGS stress tensor

$$\tau_{ij} = \overline{\rho}\left(\widetilde{u_i u_j} - \widetilde{u}_i \widetilde{u}_j\right), \tag{2.7}$$

the SGS heat flux

$$Q_j = \overline{\rho}\left(\widetilde{u_j T} - \widetilde{u}_j \widetilde{T}\right), \tag{2.8}$$

and the SGS turbulent diffusion

$$J_j = \frac{1}{2}\left(\overline{\rho u_j u_i u_i} - \overline{\rho}\widetilde{u}_j \widetilde{u_i u_i}\right), \tag{2.9}$$

It is suggested that the SGS turbulent diffusion can be approximated as $J_j = \tau_{ij}\widetilde{u}_j$ (Martin et al. 2000). The SGS stress tensor $\tau_{ij}$ and the SGS heat flux $Q_j$ need to be modelled based on the resolved quantities. Models for these terms are discussed below.

### 2.2. Subgrid-scale model

In LES, the structural model and eddy viscosity model are the commonly used SGS stress model for $\tau_{ij}$. The essence of the structural model is to reconstruct the SGS model directly using the resolved field, and the model is established without any prior knowledge of the interaction between the SGS and the resolved field. The representative models of the structural model are scale-similarity model (Bardina et al. 1980; Liu et al. 1994) and tensor-diffusivity model (TD model) (Clark et al. 1979; Stolz et al. 2001). The scale-similarity model is constructed on the basis of the scale similarity hypothesis, and a typical scale-similarity model is the Bardina's model, which can be expressed as

$$\tau_{ij} = \overline{\rho}\left(\widetilde{\widetilde{u}_i \widetilde{u}_j} - \widetilde{\widetilde{u}}_i \widetilde{\widetilde{u}}_j\right), \tag{2.10}$$

Nevertheless, the scale-similarity model requires test filtering in LES, which is difficult to apply in the simulation of wall turbulence with complex geometric boundary. The tensor-diffusivity model is given by

$$\tau_{ij} = \frac{1}{12}\Delta_k{}^2 \overline{\rho}\left(\frac{\partial \widetilde{u}_i}{\partial x_k} \frac{\partial \widetilde{u}_j}{\partial x_k}\right), \tag{2.11}$$

where (2.11) is obtained by a Taylor expansion, and can be interpreted as a truncation of the approximate deconvolution model for a Gaussian filter. $\Delta_k$ is the filter width (or mesh size) in $x_k$ direction. Unlike the scale-similarity model, the TD model does not require the test filtering, which may be convenient in large eddy simulation of complex turbulence.

† Email address for correspondence: cpyu@imech.ac.cn

At present, the most commonly used SGS model is the eddy-viscosity model, which is a phenomenological model. The SGS stress tensor is modelled by a term with a structure similar to viscous stress. Using a subgrid viscosity $\mu_{sgs}$ replace the molecular viscosity, the formulation of the SGS stress $\tau_{ij}$ is written as

$$\tau_{ij} - \frac{1}{3}\delta_{ij}\tau_{kk} = -2\mu_{sgs}\left(\widetilde{S}_{ij} - \frac{1}{3}\delta_{ij}\widetilde{S}_{kk}\right), \tag{2.12}$$

The most famous expression of $\mu_{sgs}$ in (2.12) is the Smagorinsky model (Smagorinsky 1963), which is obtained from the resolved strain rate tensor:

$$\mu_{sgs} = \widetilde{\rho}C_m\Delta^2\left|\widetilde{S}\right|, \tag{2.13}$$

with

$$\left|\widetilde{S}\right| = \sqrt{2\widetilde{S}_{ij}\widetilde{S}_{ij}}, \tag{2.14}$$

and $C_m$ is the coefficient of the Smagorinsky model.

Vreman (2004) proposed a new eddy-viscosity model (Vreman) by algebraic theory and it is suitable to apply in shear flows. The Vreman model is defined as

$$\mu_{sgs} = \overline{\rho}C_v\sqrt{\frac{B_\beta}{\widetilde{\alpha}_{ij}\widetilde{\alpha}_{ij}}}, \tag{2.15}$$

$$B_\beta = \beta_{11}\beta_{22} - \beta^2_{12} + \beta_{11}\beta_{33} - \beta^2_{13} + \beta_{22}\beta_{33} - \beta^2_{23}, \tag{2.16}$$

$$\beta_{ij} = \sum_{m=1}^{3}\Delta^2\widetilde{\alpha}_{mi}\widetilde{\alpha}_{mj}, \tag{2.17}$$

$$\widetilde{\alpha}_{ij} = \frac{\partial\widetilde{u}_j}{\partial x_i}, \tag{2.18}$$

The model coefficient $C_v$ is related to the Smagorinsky model coefficient $C_m$ by $C_v \approx 2.5C_m$. From previous research, we know that the structural models have high correlation with the real SGS stress but poor robustness in actual numerical simulation (Horiuti 1989). On the contrary, the eddy-viscosity models have high robustness but weak correlation with the real SGS stress (Garnier *et al.* 2009; Yu *et al.* 2017).

## 3. Energy flux similar method

Since the proposal of first concept of energy cascade by Richardson, the research on energy cascade have been always the core content of turbulence studies (Pope 2000). In 1941, Kolmogorov formulated the energy cascade for the first time (Kolmogorov 1941) and suggested that the energy flux is constant in the inertial subrange of locally isotropic turbulence, where the energy flux refers to the energy transfer rate from the large scale to small scale. Subsequently, increasingly more research have been focused on the kinetic energy flux of compressible and incompressible turbulent flows (Meneveau & Sreenivasan 1987; Borue & Orszag 1998; Eyink 2006; Wang et al. 2013). The kinetic energy flux between different-scale eddies is the essence of energy cascade and it reflects the dynamic process of the generation and evolution of turbulence. Accurate prediction of energy flux at any scale is the guarantee of simulating turbulent flows accurately.

At the given scale Δ, the filtered kinetic energy equation can be written as

$$\frac{\partial\left(\frac{1}{2}\overline{\rho}\widetilde{u}_i^{\ 2}\right)}{\partial t} + \frac{\partial}{\partial x_j}\mathbf{J}_\Delta = \Pi_\Delta + D_\Delta + \overline{p}\frac{\partial\widetilde{u}_i}{\partial x_i}, \tag{3.1}$$

$$\mathbf{J}_\Delta = \frac{1}{2}\overline{\rho}\widetilde{u}_i\widetilde{u}_j - \widetilde{u}_i\widetilde{\sigma}_{ij} + \widetilde{u}_i\tau_{ij} + \overline{p}\widetilde{u}_j, \tag{3.2}$$

$$D_\Delta = \widetilde{\sigma}_{ij}\frac{\partial\widetilde{u}_i}{\partial x_j}, \tag{3.3}$$

$$\Pi_\Delta = \tau_{ij}\frac{\partial\widetilde{u}_i}{\partial x_j} \tag{3.4}$$

where $J_\Delta$ is spatial transport of large-scale kinetic energy, $\overline{p}\frac{\partial\widetilde{u}_i}{\partial x_i}$ is large-scale pressure dilatation, $D_\Delta$ is the viscous dissipation acting on the large scale. Here, $\Pi_\Delta$ is the kinetic energy flux term from scale Δ to the smaller scale, and it can be also called SGS dissipation.

In transitional and turbulent flows, on the given mesh scales Δ, the total dissipation $\varepsilon_\Delta$ should be expressed as

$$\varepsilon_\Delta = D_\Delta + \Pi_\Delta, \tag{3.5}$$

For the transitional flow, the laminar flow and turbulence coexist with irregular spatial and temporal distributions, which is the phenomenon of spatiotemporal intermittency (Chat´e & Manneville 1987; Tritton 2012). The energy flux $\Pi_\Delta$ through the mesh scale Δ is approximately equal to zero in the laminar region, and it cannot be ignored in the turbulent region. While in the full turbulence, we deem that the local energy flux exist in the whole region of the turbulent flow. In the filtered kinetic energy equation 3.1, the energy flux $\Pi_\Delta$ is an unclosed term and needs to be modelled. Based on a tensor eddy viscosity, Borue & Orszag (1998) suggested a simple parametrization for the local energy flux in the inertial subrange of homogeneous and isotropic turbulence for the first time. Subsequently, Eyink (2006) developed a multi-scale gradient expansion of energy flux in the incompressible homogeneous turbulence.

(3.4) shows that the energy flux is proportional to the product of the SGS stress tensor and the resolved velocity gradient tensor. In order to obtain the proper energy flux similar to the real energy flux in complex turbulent flows, a suitable SGS stress model τij should be selected. Furthermore, we select the TD model, Smagorinsky model and Vreman model from the two categories of commonly used SGS models to perform a priori tests of the modelled energy flux.

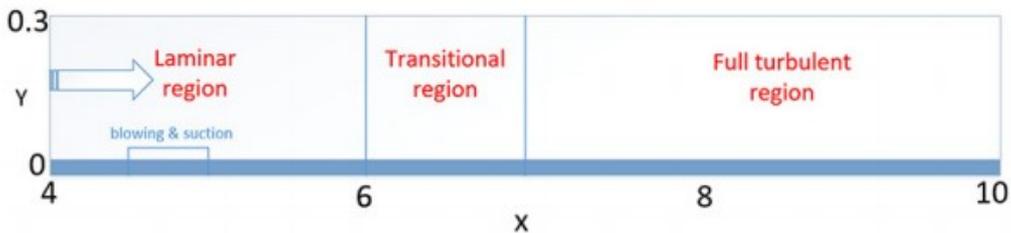

Figure 1. Sketch of the computational domain for the numerical simulation

† Email address for correspondence: cpyu@imech.ac.cn

First, we use direct numerical simulation (DNS) data of a spatially developing supersonic adiabatic flat plate boundary layer flow (Pirozzoli et al. 2004) (at $Ma = 2.25$ and $Re_\theta \approx 4000$) to analyse the similarity between the modelled energy fluxes with the real energy flux. The grid resolution for DNS is $10090 \times 90 \times 320$, and the DNS employs a seventh-order difference scheme for spatial discretization and a third-order Runge-Kutta method for time advancement. The 'viscous derivatives' and viscous flux function are determined by a sixth-order difference scheme. The computational domain (see Figure 1) is bounded by in-flow and out-flow boundaries, a wall boundary, a far-field boundary, and the two boundaries (periodic) in the span-wise direction. The size of the computational domain is $L_x \times L_y \times L_z = 6 \times 0.3 \times 0.175$ and $\Delta x^+ \times \Delta y^+ \times \Delta z^+ = 6.02 \times 0.58 \times 5.47$ in the stream-wise, wall-normal, and span-wise directions.

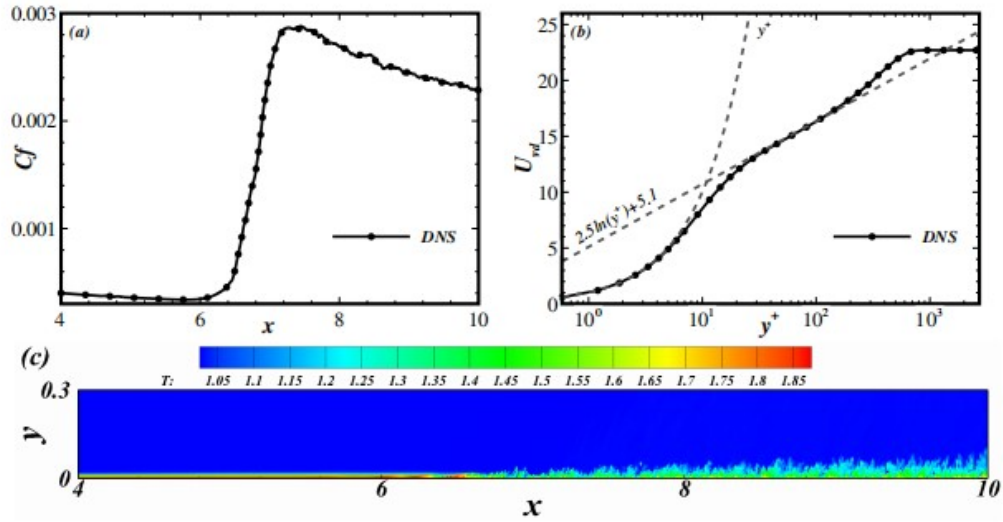

Figure 2. DNS results of the flat-plate flow. (a) Distribution of skin friction; (b) Distribution of the van-Driest transformed mean stream-wise velocity at x = 8.8; (c) The instantaneous temperature field.

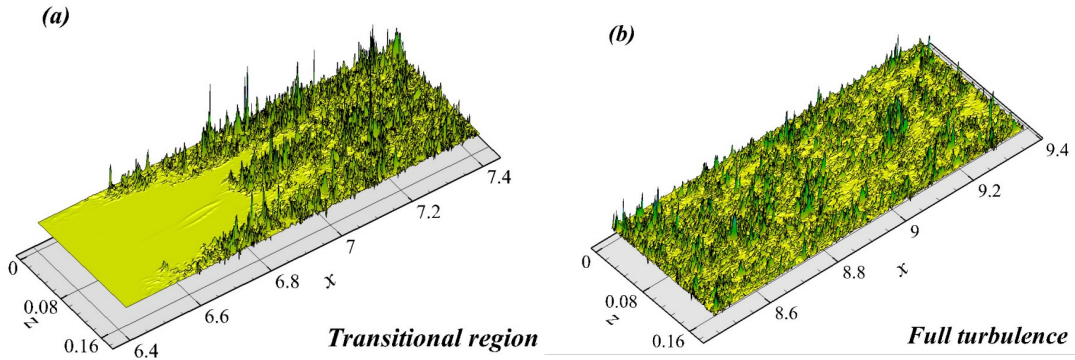

Figure 3. Distribution of instantaneous real energy flux field in transitional region and full turbulent region at $y^+ = 15$.

Figure 2 shows the DNS results of the flat-plate flow, and the profile of skin friction (Figure2(a)) reveals the regions of laminar flow at $4 \le x \le 6$, transitional flow at $6 \le x \le 7.5$, and fully developed turbulence at $x > 7.5$. Figure2(b) shows the mean velocity at x = 8.8. The result of DNS are in good agreement with the theoretical solution. The visual distribution of temperature flow field is presented in Figure2(c).

In a priori tests, the DNS data are filtered in the span-wise direction with the top-hat filter; the filter width is $\overline{\Delta} = 8\Delta_z$. Using these data, the real energy flux and energy flux of different models across the scale $\overline{\Delta}$ can be obtained. Figure 3 shows the distribution of the real local energy flux field at $y^+ = 15$. Energy flux is evenly distributed in the region of full turbulence. On the contrary, energy flux is highly intermittent in the transitional region in terms of amplitude and distribution of space. The transition appears to first occur on both sides of the flat-plate, and the amplitude of energy flux at transition peak is the largest.

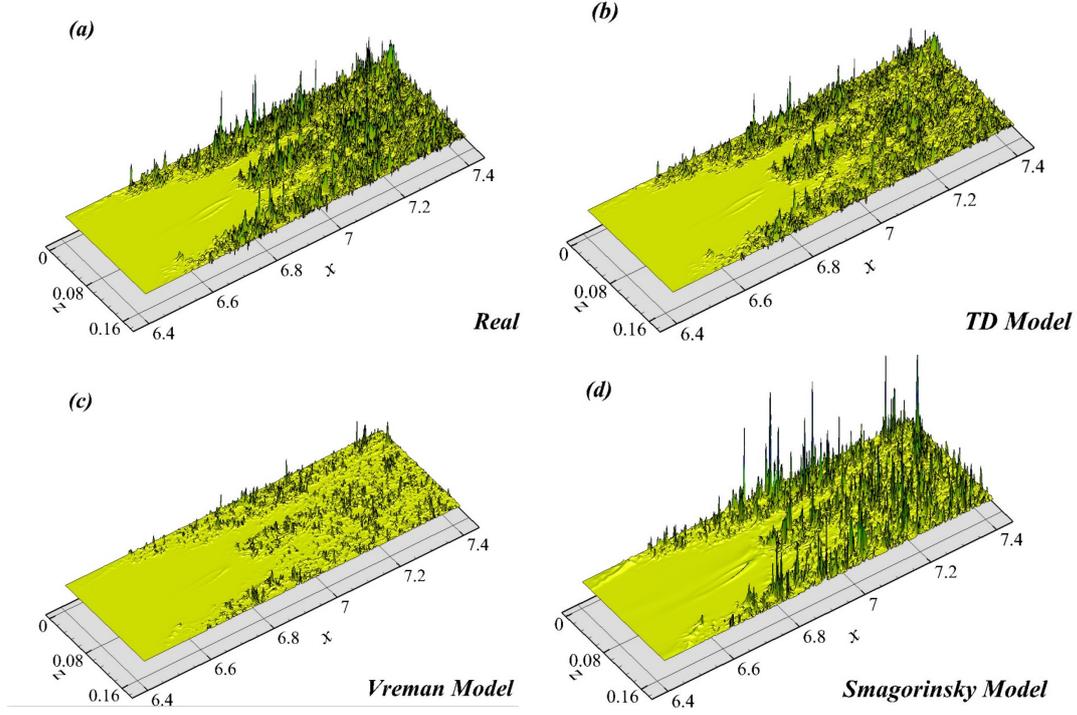

Figure 4. Distribution of modelled local energy fluxes contrast with the real local energy flux in transitional region at y+ = 15. (a) The real local energy flux; (b) The local energy flux from the TD model; (c) The local energy flux from the Vreman model; (d) The local energy flux from the Smagorinsky model.

Figure 4 shows the distribution of the modelled energy flux in the transitional region at $y^+ = 15$. The energy flux obtained from the TD model has a similar distribution with the real energy flux, with only a slight deviation in amplitude. Furthermore, the distributions of energy flux calculated using the Vreman and Smagorinsky models significantly differ from that of the real energy flux. Simultaneously, we quantitatively analyse the correlation between the modelled energy fluxes and the real energy flux using the correlation coefficient γ, for which the expression is

$$\gamma = \frac{\left\langle \left(M - \langle M \rangle\right)\left(R - \langle R \rangle\right)\right\rangle}{\left(\left\langle \left(M - \langle M \rangle\right)^2 \right\rangle \left\langle \left(R - \langle R \rangle\right)^2 \right\rangle\right)^{1/2}}, \tag{3.6}$$

where <•> denotes the ensemble average, which can be regarded as the spatial average along the span-wise direction of the flat plate, M denotes the modelled energy flux, and R denotes the real energy flux.

Figure 5 shows the correlation coefficients γ between the real energy flux and the modelled energy flux for different cases obtained a priori using DNS data of flat plate flow. Figure 5 (a, b, c) displays the correlation coefficients γ along the stream-wise direction at three different normalized normal heights $y^+$ and the fixed filter width $\overline{\Delta}$ ( $\overline{\Delta} = \Delta_z$ ). In the laminar region ($4 \leq x \leq 6$), the real energy flux and energy fluxes from the TD model and Vreman model are approximate to zero, and the energy flux from the Smagorinsky model has a non-negligible value. Therefore, we consider the correlation coefficients γ from both TD and Vreman models to be 1 and γ from Smagorinsky model to be 0. In the transitional region ($6 \leq x \leq 7$), the correlation coefficient from the TD model is close to or over 0.9. However, the γ obtained from the Vreman model sharply declines from 1 to less than 0.5, and that from the Smagorinsky model remains very low.

† Email address for correspondence: cpyu@imech.ac.cn

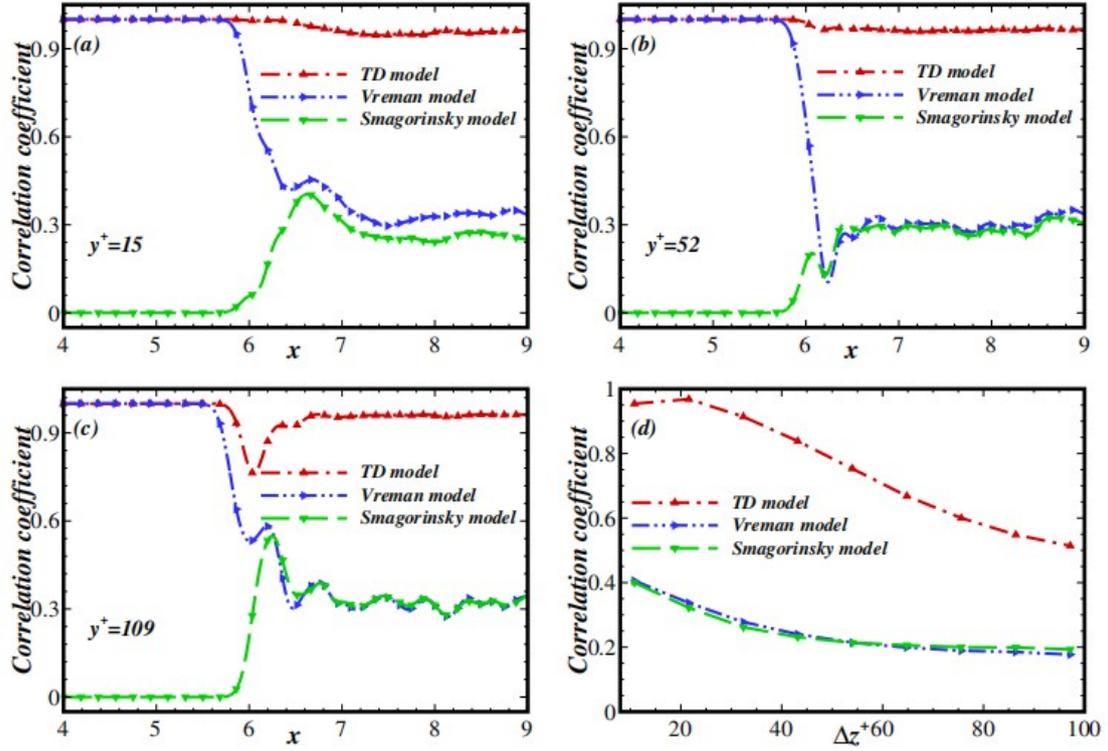

Figure 5. Distribution of correlation coefficient along the stream-wise direction at different $y^+$ under $8\Delta_z$ filter width. (a) $y^+$= 15; (b) $y^+$= 52; (c) $y^+$= 109; (d) The variation of correlation coefficient under different filter scales at x = 8.8.

In the turbulent region ($7 \leq x \leq 9$), the correlation coefficient γ from the TD model remains above 0.9 and the γ from both the Vreman and Smagorinsky models remain below 0.4. In addition, figure 5(d) shows the variation of the correlation coefficient with respect to the normalized filter width at the settled location (x = 8.8, $y^+$ = 15). The correlation coefficients from all models decrease with increasing filter width, but the energy flux from the TD model continues to have a much higher correlation with the real energy flux than those from the eddy viscosity models. Seeing from figure 4 and qualitative analysis in figure 5, we could infer that the energy flux from the TD model and the real energy flux have a perfect structural correlation, except for a slight difference in amplitude. In view of this cognition, using a priori results of the flat-plate flow at $y^+$ = 15, we apply a simple modification to the energy flux from the TD model through the multiplication of fixed coefficient 1.3 and get the modified energy flux. Figure 6 shows the results of the comparison between the modified modelled energy flux and the real energy flux, which exhibit

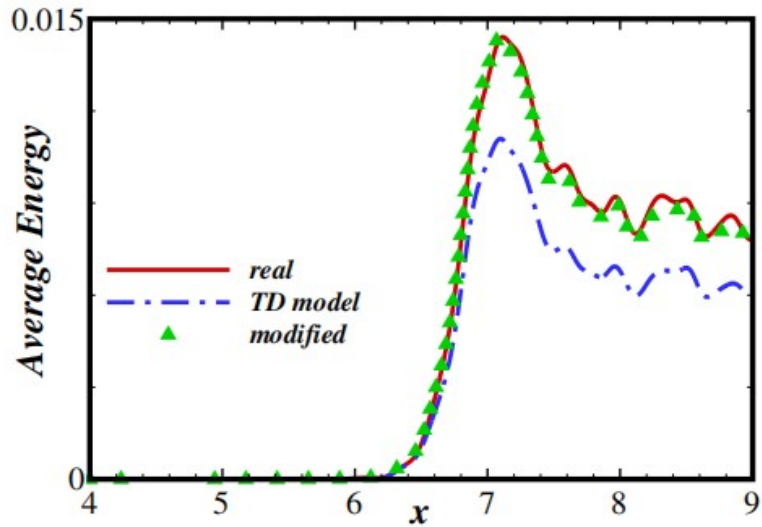

Figure 6. Ensemble average of the real energy flux, the energy flux from the TD model and the modified energy flux from the TD model under $8\Delta_z$ filter width at $y^+= 15$.

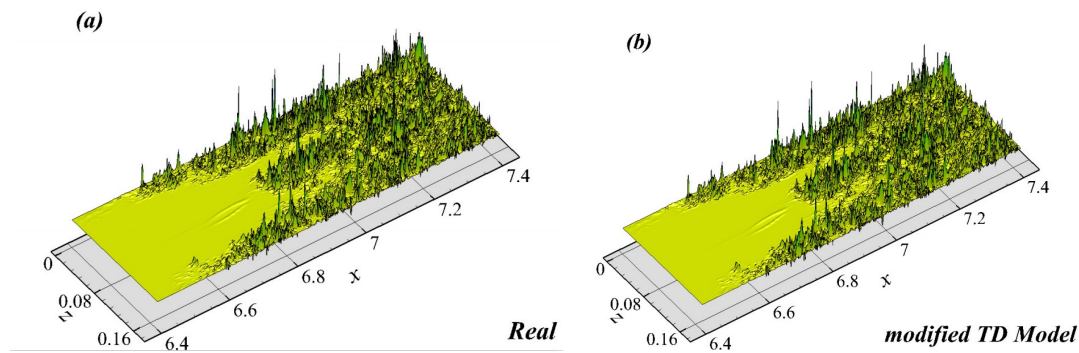

Figure 7. Distribution of the local real energy flflux and modifified energy flflux from the TD model in transitional region at $y^+= 15$.

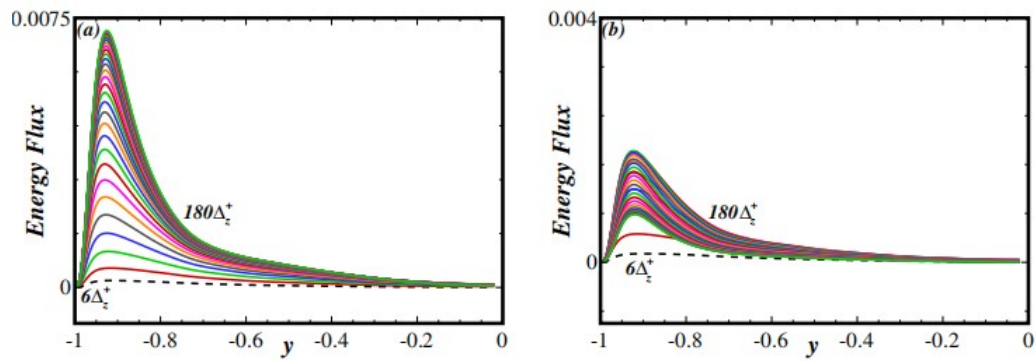

Figure 8. Distribution of energy flux along the normal direction in different filter scales. (a) The real energy flux; (b) The energy flux from the TD model. The horizontal axis represents the normal height of the half channel (-1 is the wall of the channel, 0 is the center line of the channel), and the vertical axis represents the magnitude of the energy fluxes.

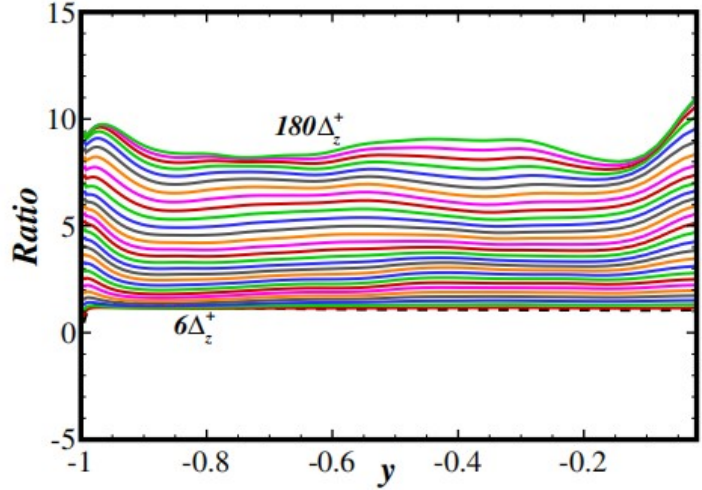

Figure 9. The ratio of the real energy flux to the energy flux from the TD model along the normal direction in different filter scales.

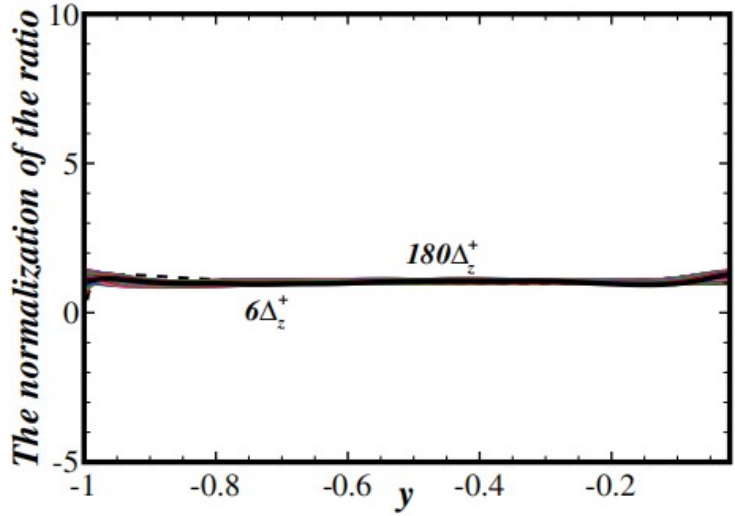

Figure 10. The normalized ratio with the formulation $\eta_\Delta$.

perfect agreement. To further verify the modified results, the distributions of the local real energy flux and the modified energy flux from the TD model at $y^+ = 15$ were determined. Figure 7 shows that the structures of the real energy flux and the modified energy flux are almost the same in terms of amplitude and distribution of space. To ensure highly similarity between the modified energy flux and the real energy flux at different locations and filter widths, we rectify precisely the energy flux from the TD model using the DNS data of a temporally turbulent compressible channel flow (at $Ma = 1.5$). The computation domain for the DNS of the channel flow is a box with a size of $4\pi \times 2 \times 4/3\pi$, and the grids for the DNS are $900 \times 201 \times 300$ and $\Delta x^+ \times \Delta y^+_{wall} \times \Delta z^+ = 3 \times 0.32 \times 3$. In a priori tests, the DNS data are filtered in the stream-wise and span-wise directions with the top-hat filter. Figure 8 shows the distribution of the energy flux across different filter widths along the normal direction, and figure 8 (a) shows the results of the real energy flux and the energy flux from the TD model in figure 8 (b). As shown in figure 8(a) and figure 8(b), both the real energy flux and the modelled energy flux obey the similar regulation that the maximum energy flux occurs near the buffer region and tend to 0 near the centre line, and the energy fluxes also present a certain regularity with increasing filter width. Figure 9 shows the ratio

| | Grids | $\Delta x^+$ | $\Delta y_w^+$ | $\Delta z^+$ |
|---|---|---|---|---|
| DNS | $900 \times 201 \times 300$ | 2.99 | 0.32 | 2.99 |
| Smagorinsky | $48 \times 65 \times 48$ | 54.32 | 1.01 | 18.11 |
| Vreman | $48 \times 65 \times 48$ | 56.48 | 1.05 | 18.83 |
| WALE | $48 \times 65 \times 48$ | 54.86 | 1.02 | 18.28 |
| DSM | $48 \times 65 \times 48$ | 57.55 | 1.07 | 19.18 |
| EFSM(new method) | $48 \times 65 \times 48$ | 57.55 | 1.07 | 19.18 |

TABLE 1. Parameters of the simulations in channel flow ($Ma = 1.5$, $Re = 3000$).

of the real energy flux to the energy flux from the TD model across different filter scales along the normal direction. In figure 9, the ratio does not exhibit any apparent change along the normal direction, but it changes significantly with varying filter widths. From the analysis results, we attempt to fit out the formulation $\eta_\Delta$ of the ratio and the normalized filter width $\Delta/\Delta^+_w$, where

$$\Delta^+_w = \frac{\rho_\omega u_\tau}{\mu_w},\qquad(3.8)$$

here $u_\tau = \sqrt{\tau_w / \rho_w}$ is wall friction velocity and $\tau_w = \mu_w \dfrac{\partial u}{\partial y}$ is wall shear stress. The

formulation of ratio can be expressed as

$$\eta_\Delta = C_1 \left( \frac{\Delta}{\Delta^+_w} \right)^2 + C_2 \left( \frac{\Delta}{\Delta^+_w} \right) + 1.0,\qquad(3.8)$$

where $C_1 \approx 8 \times 10^{-5}$, $C_2 \approx 0.01$.

Using the formulation (3.8), the new ratio of the real energy flux to the modified energy flux from the TD model is approximate to 1 at different locations and filter widths, as shown in figure 10.

Previous research found that the TD model has poor robustness in LES due to the lack of SGS dissipation. In this study, we found that under the same SGS dissipation, the TD model still lack of robustness compared to the eddy-viscosity model.

In order to secure the robustness of computation and the simulation results similar with the real flow, we develop the energy flux similarity method (EFSM) in LES of transitional and turbulent flows. As a typical eddy-viscosity model, the Smagorinsky model is selected in the actual computation. The Smagorinsky model can be expressed as

$$\tau_{ij}^{\ sm} = -2C_{sm}\overline{\rho}\Delta^2 \left| \widetilde{S} \right| \left( \widetilde{S}_{ij} - \frac{1}{3}\delta_{ij}\widetilde{S}_{kk} \right) + \frac{2}{3}C_1\overline{\rho}\Delta^2 \left| \widetilde{S} \right|^2 \delta_{ij},\qquad(3.9)$$

For energy-flux similarity criterion, the following is required

$$\Pi^{mtd} = \Pi^{sm},\qquad(3.10)$$

$$\Pi^{sm} = \Pi_1^{\ sm} + \Pi_2^{\ sm},\qquad(3.11)$$

and the two terms of the energy flux from the Smagorinsky model can be written as

$$\Pi_1^{\ sm} = -2C_{sm}\overline{\rho}\Delta^2 \left| \widetilde{S} \right| \left( \widetilde{S}_{ij} - \frac{1}{3}\delta_{ij}\widetilde{S}_{kk} \right),\qquad(3.12)$$

$$\Pi_2{}^{sm} = \frac{2}{3} C_1 \overline{\rho} \Delta^2 \left| \widetilde{S} \right|^2 \delta_{ij} \widetilde{S}_{ij}, \tag{3.9}$$

Furthermore, the modified energy flux from TD model must be divided into two parts $\Pi^{mtd}{}_1$ and $\Pi^{mtd}{}_2$ as

$$\Pi_1{}^{mtd} = \eta_\Delta \left( \tau_{ij}{}^{td} - 1/3 \tau_{kk}{}^{td} \delta_{ij} \right) \widetilde{S}_{ij}, \tag{3.14}$$

$$\Pi_2{}^{mtd} = 1/3 \eta_\Delta \tau_{kk}{}^{td} \delta_{ij} \widetilde{S}_{ij}, \tag{3.15}$$

Then, we let $\Pi^{mtd}{}_1 = \Pi^{sm}{}_1$ and $\Pi^{mtd}{}_2 = \Pi^{sm}{}_2$, and the coefficients of the Smagorinsky model can be presented as

$$C_{sm} = -\eta_\Delta \frac{\left( \tau_{ij}{}^{td} - 1/3 \tau_{kk}{}^{td} \delta_{ij} \right) \widetilde{S}_{ij}}{2 \overline{\rho} \Delta^2 \left| \widetilde{S} \right|^2 \delta_{ij} \widetilde{S}_{ij}}, \tag{3.16}$$

$$C_1 = \eta_\Delta \frac{\tau_{kk}{}^{td} \delta_{ij} \widetilde{S}_{ij}}{2 \rho \Delta^2 \left| \widetilde{S} \right|^2 \delta_{ij} \widetilde{S}_{ij}}, \tag{3.17}$$

For most transitional and turbulent flows, the Smagorinsky model can be written as

$$\tau_{ij}{}^{sm} = -2 C_{sm} \overline{\rho} \Delta^2 \left| \widetilde{S} \right| \left( \widetilde{S}_{ij} - \frac{1}{3} \delta_{ij} \widetilde{S}_{kk} \right), \tag{3.18}$$

From (3.10), the coefficient of the Smagorinsky model can be confirmed as

$$C_{sm} = -\eta_\Delta \frac{1/12 \Delta_k{}^2 \overline{\rho} \left( \dfrac{\partial \widetilde{u}_i}{\partial x_k} \dfrac{\partial \widetilde{u}_j}{\partial x_k} \right) \widetilde{S}_{ij}}{2 \rho \Delta^2 \left| \widetilde{S} \right| \left( \widetilde{S}_{ij} - \dfrac{1}{3} \delta_{ij} \widetilde{S}_{kk} \right) \widetilde{S}_{ij}}, \tag{3.19}$$

## 4. Application

### 4.1. Energy flux similarity method for LES of turbulent channel flow

In this section, the energy flux similarity method is applied to the LES of the compressible turbulent channel flows. The case setting of LES is same as the DNS in §3. The filtered Navier-Stokes equations (2.1)-(2.3) are solved using a finite difference solver in cartesian coordinates, the equations are temporally integrated using the third-order R-K scheme and a sixth-order central difference scheme is used for the discretization of both the convective and viscous terms. Details of the grids are listed in Table 1. In this simulation, the following SGS models are selected for comparison:

1. constant coefficient Smagorinsky model (SM),
2. constant coefficient Vreman model (Vreman),
3. Wall-Adapting Local Eddy-viscosity model (WALE),
4. dynamic Smagorinsky model (DSM),

The distribution of the Van-Driest transformed mean stream-wise velocity ( $U_{vd} = \int_0^U \sqrt{\langle \rho \rangle / \rho_w} \, d\langle U \rangle$ ) as a function of $y^+$ is displayed in Figure 11. As expected for $y^+ < 5$, the velocity evolves linearly with $y^+$. Meanwhile, all the results of the SGS models collapse to the DNS result in the viscous sub-layer and up to the buffer region ($y^+ < 25$). In the log-law region, the EFSM provides a perfect

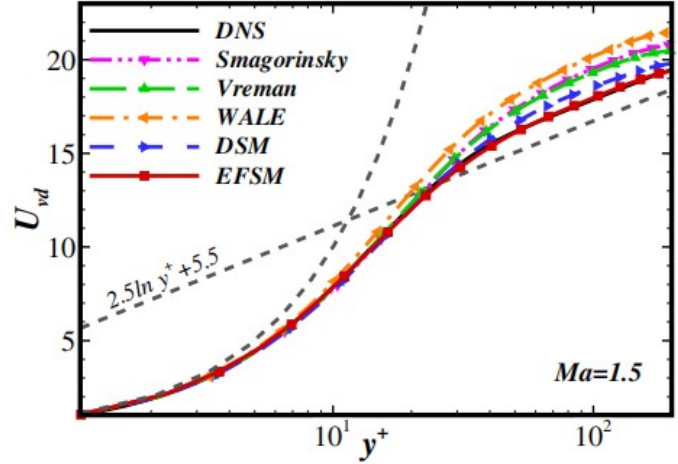

Figure 11. Distribution of the van-Driest transformed mean stream-wise velocity from DNS and different SGS models.

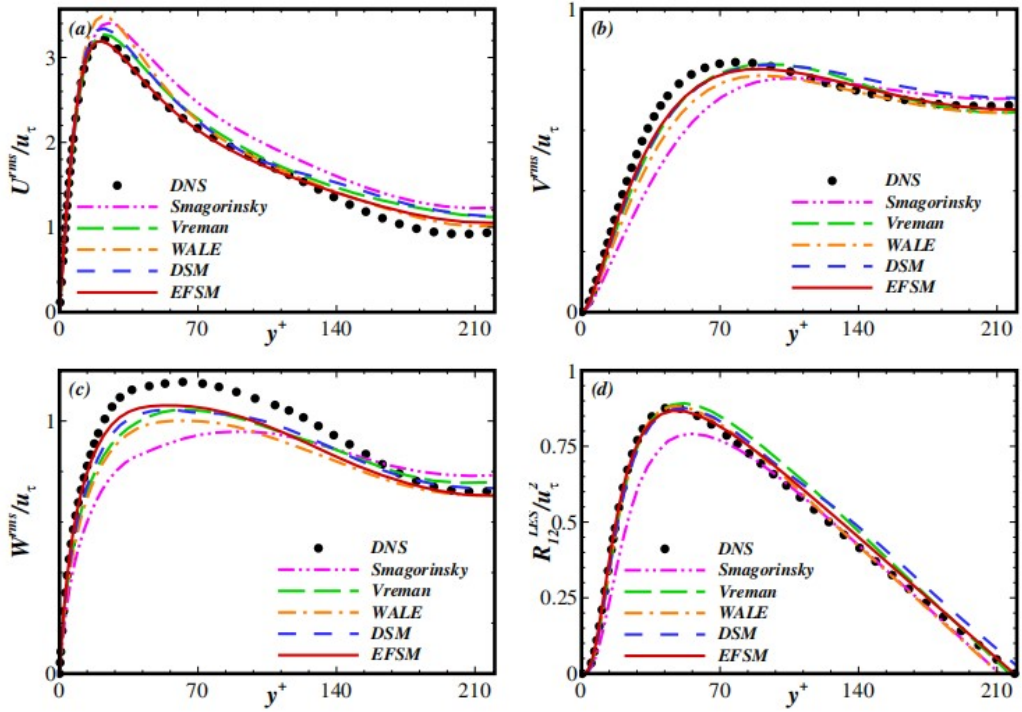

Figure 12. Profiles of turbulence intensities and resolved Reynolds stress normalized by friction velocity $u_\tau$ versus $y^+$: (a) stream-wise turbulence intensity $U^{rms}$;(b) normal-wise turbulence intensity $V^{rms}$;(c) span-wise turbulence intensity $W^{rms}$;(d) the resolved Reynolds stress $R^{LES}_{12}$ .

estimation of $U_{vd}$, but the results of other SGS models show an obvious deviation from the DNS result. Furthermore, EFSM provides proper SGS dissipation, and all the other SGS models support excessive SGS dissipation. From the analysis, we know that the WALE model is the most dissipative, the Vreman and Smagorinsky models have the same dissipation, and the DSM still overestimates dissipation. Figure 12 (a, b, c) shows the profiles of the resolved turbulence intensities $\widetilde{u}_i^{\,rms} = \left\langle \left( \widetilde{u}_i - \langle \widetilde{u}_i \rangle \right)^2 \right\rangle^{1/2}$ obtained from DNS, and several SGS models including SM, WALE, Vreman, DSM, and EFSM. Figure 12(a) shows the stream-wise turbulence intensity $U^{rms}$

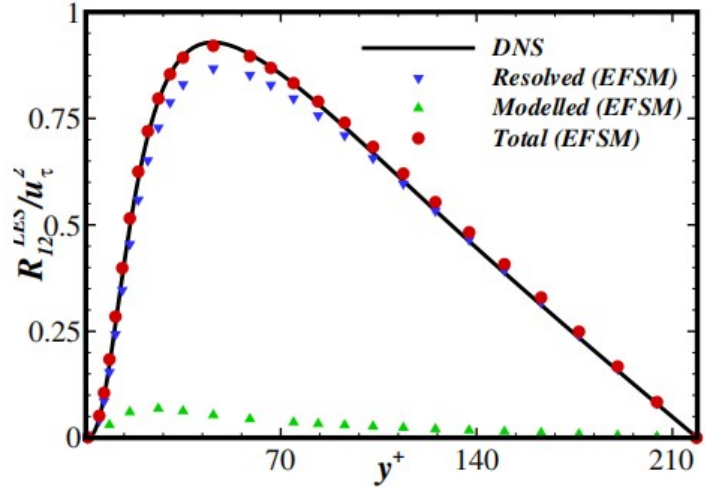

Figure 13. Relative contributions of the modelled and resolved Reynolds stresses to the total
Reynolds stress of EFSM and the total Reynolds stress of DNS.

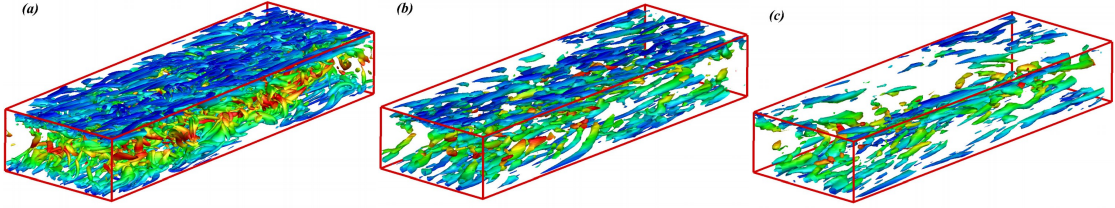

Figure 14. Instantaneous isosurface of $Q$ (second invariant of the strain rate tensor) getting
from (a) DNS, (b) EFSM, (c) WALE in turbulent channel flow.

, and the EFSM result shows good agreement with the DNS data especially in the buffer region, the Smagorinsky model shows the largest deviation from the DNS data, and both WALE and Vreman models have lower performance than EFSM. Figure 12(b, c) show the normal-wise turbulence intensity $V^{rms}$ and span-wise turbulence intensity $W^{rms}$, respectively. The two figures show that all the SGS models have low performance, but the results of EFSM are better than those of the other SGS models. Figure 12 (d) shows the resolved Reynolds stress $R^{LES}_{12}$ with the normalized normal height $y^+$. Compared with the other SGS models, EFSM simulates Reynolds stress $R^{LES}_{12}$ well, especially in the near-wall and buff regions. Although the Van-Driest damping function in the near-wall and buffer regions was adopted in the Smagorinsky model, the Smagorinsky model fails to predict the resolved Reynolds stress $R^{LES}_{12}$ . In Figure 13, we show the relative contributions of the resolved and modelled Reynolds stresses to the total Reynolds stress of EFSM contrast to the total Reynolds stress of DNS. If the the flow is assumed ergodic, the expression of the total Reynolds stress from LES results could be written as

$$R_{ij} = \left\langle u_i u_j \right\rangle - \left\langle u_i \right\rangle \left\langle u_j \right\rangle = R_{ij}^{LES} + \left\langle \tau_{ij}^{mod} \right\rangle, \qquad (4.1)$$

where the resolved Reynolds stress $R_{ij}^{LES} = \left\langle \widetilde{u}_i \widetilde{u}_j \right\rangle - \left\langle \widetilde{u}_i \right\rangle \left\langle \widetilde{u}_j \right\rangle$ . As expected, the total Reynolds stress obtained from EFSM almost coincides completely with the real Reynolds stress from DNS data.

Figure 14 shows the instantaneous isosurface of $Q$ selected from DNS, EFSM, and WALE. The EFSM results possess abundant small-scale structures and resemble the coherent structure distribution of DNS. On the contrary, the Q of WALE lacks small-scale structures and the structure distribution widely differs from those of DNS and EFSM.

| Model | Smagorinsky | Vreman | WALE | DSM | EFSM |
|-------|-------------|--------|------|-----|------|
| time | 0.033s | 0.033s | 0.033s | 0.055s | 0.036s |

TABLE 2. The average wall time per one time step for different models by using 36 CPU cores of the TIANHE-1.

| case | $N_x$ | $N_y$ | $N_z$ | $\Delta x^+$ | $\Delta y_w^+$ | $\Delta z^+$ |
|------|-------|-------|-------|--------------|----------------|--------------|
| DNS | 10090 | 90 | 320 | 6.02 | 0.58 | 5.47 |
| LES-grid1 | 1500 | 90 | 100 | 40.1 | 0.58 | 17.5 |
| LES-grid2 | 1000 | 90 | 80 | 60.2 | 0.58 | 21.9 |
| LES-grid3 | 1000 | 60 | 50 | 60.2 | 1.00 | 35.1 |

TABLE 3. Parameters of the simulations in supersonic transition and turbulent flat-plate boundary layer ($Ma = 2.25$, $Re = 635000$).

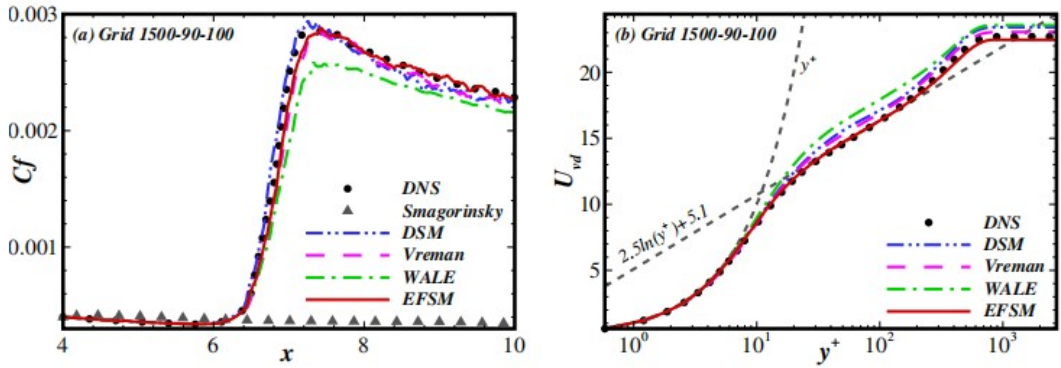

Figure 15. The LES results of Grid-1. (a) Skin friction distribution; (b) Distribution of mean stream-wise velocity at x = 8.8.

Based on the Q analysis, we could consider EFSM to have good prediction ability for local turbulent structures.

Table 2 shows the average wall time per time step from the SGS models. In terms of the non-filtering procedure, EFSM shows almost the same computational efficiency as the constant coefficient models. At the same time, the computational efficiency of EFSM is obviously higher than that of DSM, which needs test filtering

4.2. Energy flux similarity method for LES of supersonic transition and turbulent flat-plate boundary layer

In the previous section, the good performance of the new method based on energy flux similarity for fully developed wall turbulence was validated. We also examine the performance of the new method in LES of flat-plate boundary layer flow. Flat-plate boundary layer flow is a typical flow with laminar, transitional, and full turbulent regions, which can be regarded as a classical case for evaluating the performance of the new method in transitional and turbulent flows. The flow parameters are introduced in §3, and it can also be found in (Pirozzoli et al. 2004).

The grids used for LES of the flat-plate flow are shown in Table 3, and the computational domain is the same as that shown in figure 1 (§3). The mesh size in the x direction is uniform; fine mesh is adopted near the wall, and uniform mesh is adopted in the span-wise direction. Blow and suction disturbance are imposed at the wall with the interval of $4.5 = x_a \le x \le x_b = 5.0$. The form of blowing and suction is the same as that in reference (Pirozzoli et al. 2004), except for the magnitude of amplitude. In order to simulate natural transition, an amplitude of 0.02 is selected in this case.

† Email address for correspondence: cpyu@imech.ac.cn

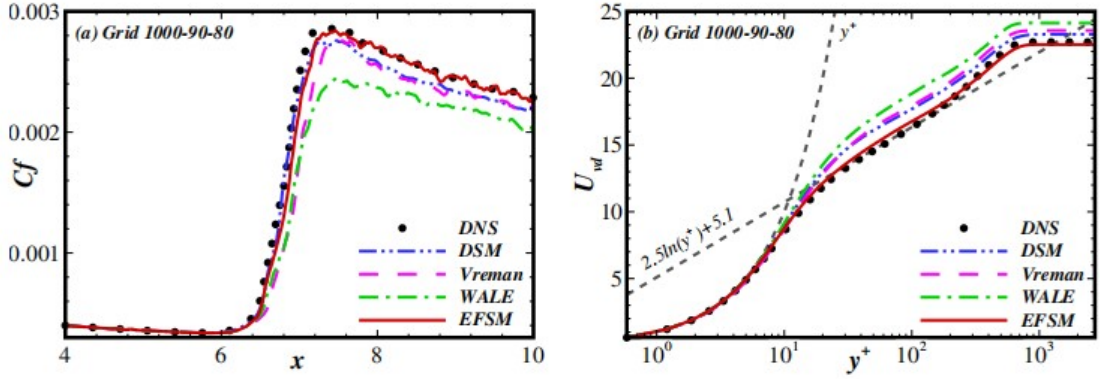

Figure 16. The LES results of Grid-2. (a) Skin friction distribution; (b) Distribution of mean stream-wise velocity at x = 8.8.

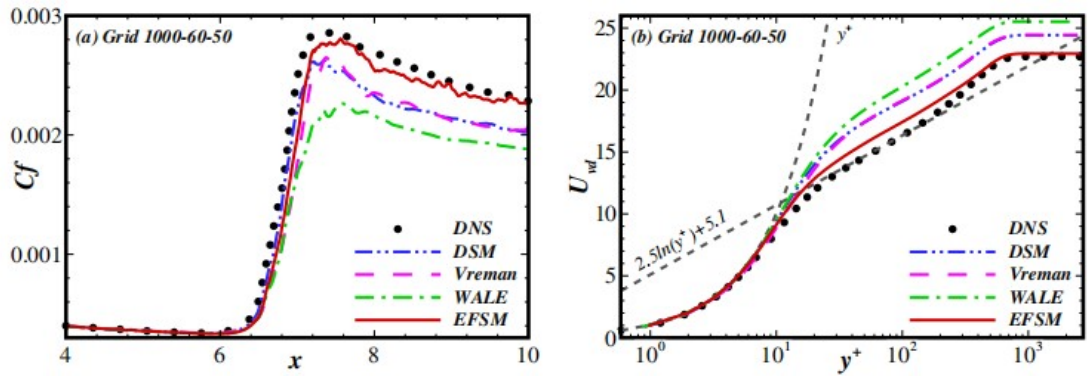

Figure 17. The LES results of Grid-3. (a) Skin friction distribution; (b) Distribution of mean stream-wise velocity at x = 8.8.

Figure 15 (a) shows the skin friction from the SGS models on Grid-1 compared to the DNS data. As clearly shown in figure 15 (a), the Smagorinsky model fails to predict the transition process. This is because the model largely overestimates dissipation, which will hinder the development of the disturbance wave. On the other hand, the improved Smagorinsky model with EFSM can well predict the transitional process including the onset of transition and transition peak. The improved Smagorinsky model could differentiate laminar, transitional, and turbulent regions automatically, and supply appropriate SGS dissipation in different regions. This method is consistent with the viewpoint that energy flux through the grid scale is zero in the laminar region, and the distribution of energy flux has spatiotemporal intermittency in the transitional process.

At the same time, DSM and Vreman can also predict the transitional process well, and the result of WALE is clearly lower than the real value. Figure 15 (b) shows the distribution of the Van-Driest transformed mean stream-wise velocity $U_{vd}$ for Grid-1 case at x = 8.8. The results show that the velocity line of EFSM tightly collapses to the line of DNS, the profiles of DSM and Vreman slightly deviate from the profile of DNS in the low-law region, and the profile of WALE obviously deviation from that of DNS.

Figure 16 shows the LES results of Grid-2, which has a coarser grid than Grid-1. Figure 16 (a)

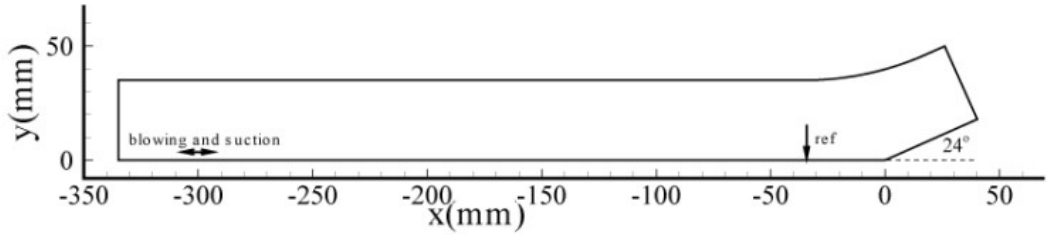

FIGURE 18. Sketch of the computational domain of compression ramp.

| case | Grid number | $\Delta x^+, x = -335$ | $\Delta x^+, x \geqslant -35$ | $\Delta y_w^+$ | $\Delta z^+$ |
|------|-------------|------------------------|-------------------------------|----------------|--------------|
| DNS | $4000 \times 160 \times 200$ | 6.52 | 2.90 | 0.58 | 4.06 |
| LES | $540 \times 100 \times 40$ | 40.60 | 40.60 | 0.87 | 20.30 |

TABLE 4. Grid parameters.

shows the skin friction along the stream-wise direction of the flat-plate flow calculated by the SGS models. In this case, EFSM still predicts the transitional process perfectly, DSM and Vreman underestimated the skin friction, and the skin-friction profiles of WALE distinctly deviate from the DNS result. The distribution of the Van□Driest transformed mean stream-wise velocity at x = 8.8 is shown in figure 16 (b); the profile predicted by EFSM is still tightly close to the DNS result. Conversely, Vreman and DSM obviously overestimate the DNS result, and WALE shows even poorer performance than in case 1.

Figure 17 shows the LES results of Grid-3, which has the coarsest grid among the three cases. Figure 17 (a) presents the skin-friction coefficient from these SGS models. In case 3, the profiles of WALE, Vreman, and DSM sharply decline from the profile of DNS including the transition peak and transition-turbulence region, but EFSM maintains better prediction results compared with the DNS results. Figure 17 (b) shows the distribution of mean streamwise velocity for case 3 at x = 8.8. The EFSM result shows a slight ascent in the log-law region but it is still evidently better than the results of WALE, Vreman, and DSM, which drastically deviate from the profile of DNS.

From figure 15,16 and 17, we find that the skin friction and mean stream-wise velocity of EFSM are perfectly consistent with the results of DNS all along. In contrast, the forecasting results of the traditional SGS models turn from good to poor with the decreasing number of grids. From the analysis, we could infer that EFSM is scale adaptive in a wider scale range.

4.3. Energy flux similarity method for LES of supersonic flow over a 24 deg compression ramp

The supersonic turbulent boundary layer over a compression ramp is a typical problem including transition, shock distortion, separation, and reattachment. It is another classical case for testing the SGS models in simulation of transition, separation, and compressible turbulence, etc. We examine the performance of EFSM in case of the supersonic turbulent boundary layer over a compression ramp.

A schematic diagram of the computation of supersonic flow over a 24 deg compression ramp is shown in figure 18; the case setting is the same as Bookey et al. (2005); Wu & Martin (2007). The computational domains are $0 \leq x \leq 35$ mm in the wall-normal direction, $0 \leq z \leq 14$ mm in the span-wise direction, and $-335 \leq x \leq 49.56$ mm in the stream-wise direction. To trigger the bypass-type transition, we impose blowing and suction perturbation on the wall at $-305 \leq x \leq -285$ mm. We provide two sets of grids: one is for DNS, and the other is for LES. The grid parameters are listed in Table4. In the upstream flat-plate region (($-335 \leq x \leq -35$ mm), the grid spacing of DNS is not uniform but a gradual process of encryption, and in the corner region(($-35 \leq x \leq 49.56$ mm), the stream-wise grid spacing is much smaller to resolve the small scales of separation flows. For the grid of LES, the stream-wise grid spacing is uniform.

† Email address for correspondence: cpyu@imech.ac.cn

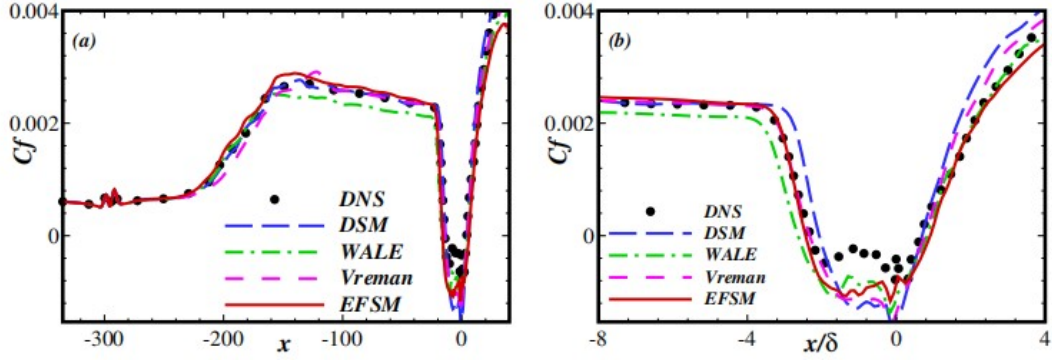

Figure 19. (a) Distribution of skin friction coefficient; (b) Distribution of skin friction coefficient in corner region, where the horizontal coordinates are normalized by the boundary-layer thickness $\delta$.

The free-stream Mach number is 2.9, free-stream Reynolds number per unit millimetre is 5581.4, and free-stream temperature is 108.1 K. Steger-Warming splitting is used for inviscid terms and then solved using the sixth-order central scheme. Viscous terms are also discretized using the sixth-order central scheme. The third-order TVD-type Runge Kutta method is used for time advancement. In order to calculate the shock wave in the corner region, we use the filter to capture the shock wave in the x ≥-35 region(Bogey *et al.* 2009).

Figure 19 shows the distribution of skin friction coefficient $C_f$ along the stream-wise direction of the flow over a compression ramp from LES and DNS. Figure 19 (a) shows the global distribution of skin friction coefficient $C_f$ in the flow. Figure 19 (a) shows that $C_f$ exhibits a drastic increase near the x =-200 mm region, denoting the occurrence of the transition. Unlike the natural transition on a flat plate in the previous case, it is the bypass transition in this case. In the corner region (($-35 \leq x \leq 35$), $C_f$ declines rapidly downstream and then reaches a negative value, indicating the occurrence of separation at this location. $C_f$ increases rapidly and shows a positive value again at x = 0 mm, indicating the reattachment of the flow. As shown in the figure, EFSM and DSM results are close to the DNS data in the whole flow field, WALE has a relatively low performance in transition-turbulence region, and Vreman predicts a deferred transition in this case.

For careful comparison of the performance of the SGS models in the separation flow, the distribution of the skin friction coefficient $C_f$ in the corner region is presented in figure 19 (b), where the horizontal coordinates are normalized by the boundary-layer thickness $\delta$. As shown in the figure, the size of the separation bubble calculated by DSM is smaller than the real size, whereas the bubble size predicted by WALE is larger. The $C_f$ predicted by Vreman in the region of reattachment slightly deviates from the real value. All the SGS models could not accurately predict the negative skin friction at the bottom of the bubble. It is encouraging that EFSM provides excellent results in both the separation and reattachment points.

According to the current turbulence theory, there are abundant backscatters in the separation region of separated turbulent flows. Figure 20 shows a partial enlarged plot of instantaneous local energy flux from LES and filtered DNS data, where the filter width of DNS is the same as the grid width of LES. From figure 20 (a), we can see that the affluent negative energy-flux regions exist in the corner of the filtered DNS flow field, which also explains that the bubble is a typical phenomenon of separated flow in this case. Figure 20 (b) also shows several blocks of negative energy-flux regions simulated by EFSM. On the contrary, Vreman cannot predict the negative energy flux displayed in figure 20 (c).

† Email address for correspondence: cpyu@imech.ac.cn

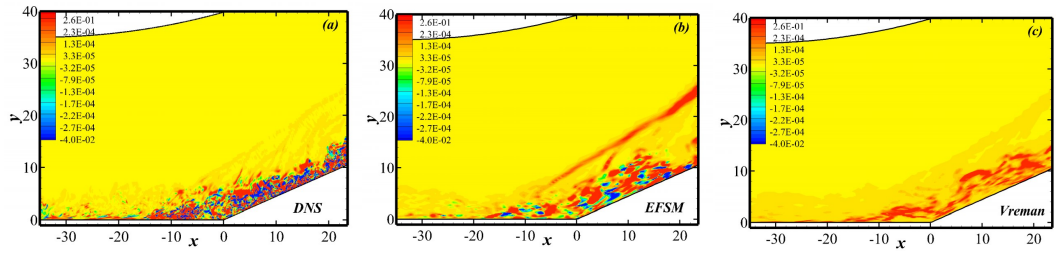

Figure 20. Partial enlarged plot in corner region of instantaneous local energy flux from LES and filtered DNS data.(a) DNS; (b) EFSM; (c) Vreman.

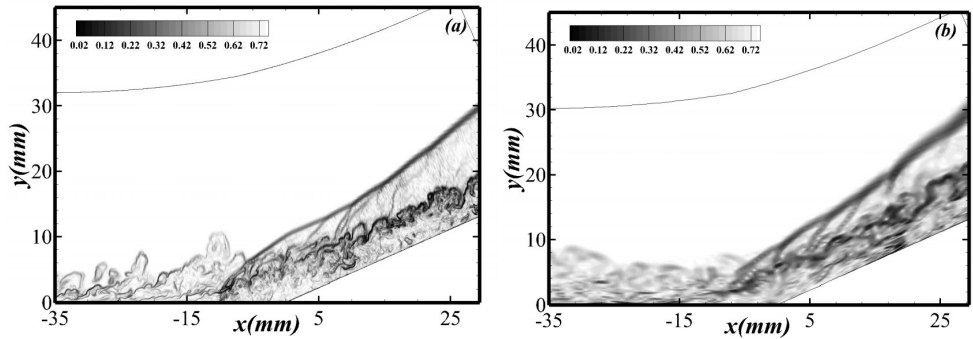

Figure 21. Numerical visualization of the instantaneous flow field in the stream-wise and wall-normal plane. (a) DNS; (b) EFSM.

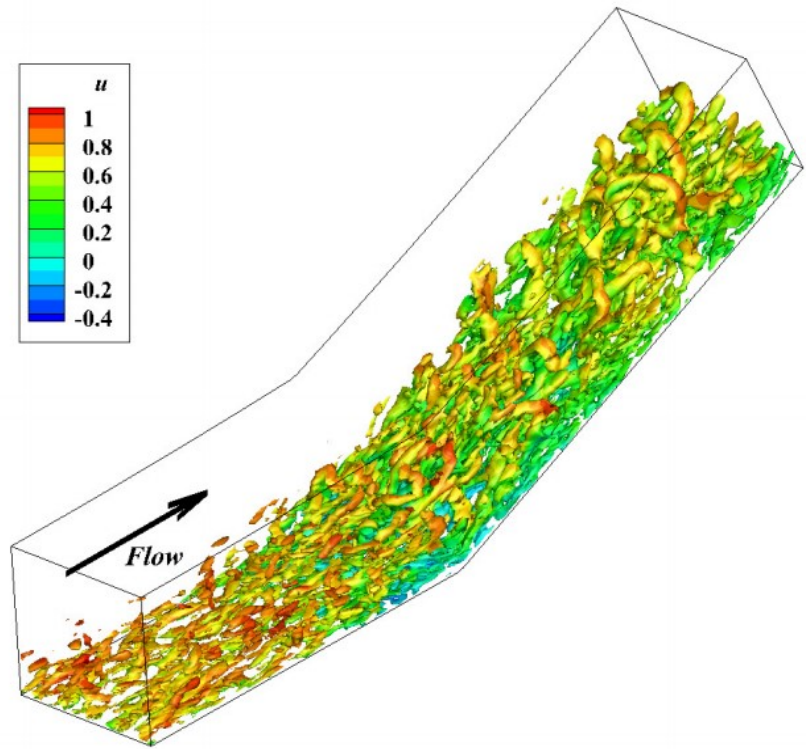

Figure 22. Isosurface of the Q criterion colored by the stream-wise velocity u.

† Email address for correspondence: cpyu@imech.ac.cn

Figure 21 presents a comparison of the numerical schlieren of the instantaneous flow field (0.8 exp [-10 ($|\Delta\rho| - |\Delta\rho|_{min}$) / ($|\Delta\rho|_{max} - |\Delta\rho|_{min}$)]) in the stream-wise and wall normal planes from DNS (figure 21 (a)) and EFSM (figure 21 (b)). The figure shows that the variation of instantaneous density gradient from LES data is similar to that from the DNS data and experimental results(Ringuette *et al.* 2009). In addition, the location of the shock wave and turbulence structure predicted by EFSM are similar to the DNS result. Coherent vortex structures obtained using the Q criterion from EFSM data are also shown in figure 22. Abundant coherent vortex structures can be observed in the flow field, such as the large-scale hairpin-like vortices.

## 5. Conclusions

We propose a new methodology based on energy flux similarity for large-eddy simulation of transitional and turbulent flows. Analysing the flow characteristics of transitional and turbulent flows, we find that the energy cascade process is the most elementary phenomenon in both transitional and turbulent flows. Nevertheless, the distribution of energy cascade in transitional flows has obvious spatiotemporal intermittency, which is the main distinguishing factor from full turbulence. According to the traditional turbulent cascade theory, the energy flux is the characteristic quantity in scale interaction of the energy cascade process, which is theoretical basis of the energy flux similarity method. Using DNS data of compressible flat-plate flow, we performed a priori tests on the correlation analysis between the real energy flux and the modelled energy flux from different SGS models, respectively. Among several SGS models, we found that the energy flux from the TD model has the strongest correlation with the real energy flux in terms of structure and distribution at different filter widths and locations of flat-plate flow, except for the amplitude of energy flux. In order to secure the amplitude of the energy flux from TD model has higher similarity with the real energy flux in different grid widths and locations of the wall turbulence, a normalized formulation $\eta_\Delta$ was applied to modify the energy flux from the TD model through a priori tests with DNS data of compressible channel flow. Thus, we obtained a uniform expression of energy flux similarity suitable for different filter widths and locations of wall turbulence. Considering the instability of the TD model in LES of turbulent flows, we selected the Smagorinsky model as the target SGS model for solving filtered N-S equations. At the same time, we also proved that the Smagorinsky model is more robust than the TD model under conditions of the same SGS dissipation. Using the energy flux similarity criterion, we obtained the new coefficient of the Smagorinsky model, finally. Our new LES method was initially applied to the simulation of compressible turbulent channel flow. Compared with other commonly used SGS models, EFSM could perfectly predict typical statistical quantities, such as mean stream-wise velocity, turbulence intensities, and Reynolds stress. Furthermore, it could describe affluent coherent structures in channel flow. We also tested the new method in supersonic spatially developing flat-plate flow. EFSM could precisely predict the natural transition process including the onset of transition and transition peak, and also provide accurate profile of skin friction and mean stream-wise velocity in cases with three different grid scales. Further, we simulated flow over a compressible ramp, and EFSM was found to accurately predict the bypass transition as well as the location of separation and reattachment in the corner region, and it could well depict the location of the shock wave and abundant coherent vortex structures.

In summary, the applicability of the new LES method has been verified with reliable physical proof. Both stability of simulation and the similarity with the real flow in LES using the energy flux similarity methodology could be confirmed. It is a scale adaptive method overcomes the scale limitation of traditional LES, and it can efficiently solve classical difficulties in LES, such as compressible flows, transitional flows, and separated flows. Moreover, the new method does not require test filtering and wall model, which makes it convenient for application to the simulation of practical flows.

## 6. Acknowledgements

This work was supported by NSFC Projects (Nos. 91852203,1472278), the National Key Research and Development Program of China (2016YFA0401200)cience Challenge Project (TZ2016001), and Strategic Priority Research Program of Chinese Academy of Sciences (Grant No. XDA17030100). The authors thank National Supercomputer Center in Tianjin (NSCC-TJ), and National Supercomputer Center in GuangZhou (NSCC-GZ) for providing computer time.

† Email address for correspondence: cpyu@imech.ac.cn

† Email address for correspondence: cpyu@imech.ac.cn